\documentclass[prl,superscriptaddress,twocolumn,preprintnumbers]{revtex4}

\usepackage{color}
\usepackage[active]{srcltx}
\usepackage{amsmath,amsfonts,amssymb,amsthm,amstext,amscd,eucal,srcltx}
\usepackage{epsfig,graphicx,bm}
\usepackage{epstopdf, epsf}
\usepackage{dcolumn}
\usepackage{hyperref}

\newcommand{\be}{\begin{equation}}
\newcommand{\ee}{\end{equation}}

\newcommand{\bse}{\begin{subequations}}
\newcommand{\ese}{\end{subequations}}
\newcommand{\bea}{\begin{eqnarray}}
\newcommand{\eea}{\end{eqnarray}}
\newcommand{\ba}{\begin{array}}
\newcommand{\ea}{\end{array}}
\newcommand{\bc}{\begin{center}}
\newcommand{\ec}{\end{center}}

\begin{document}

\pagestyle{plain}

\title{Emergent inflation from a Nambu--Jona-Lasinio mechanism in gravity with non-dynamical torsion}

\author{Andrea Addazi}
\email{andrea.addazi@lngs.infn.it}
\affiliation{Department of Physics \& Center for Field Theory and Particle Physics, Fudan University, 200433 Shanghai, China}

\author{Pisin Chen}
\email{pisinchen@phys.ntu.edu.tw}
\affiliation{Leung Center for Cosmology and Particle Astrophysics, National Taiwan University, Taipei, Taiwan 10617}
\affiliation{Department of Physics, National Taiwan University, Taipei, Taiwan 10617}
\affiliation{Kavli Institute for Particle Astrophysics and Cosmology, SLAC National Accelerator Laboratory, Stanford University, Stanford, CA 94305, U.S.A.}

\author{Antonino Marcian\`o}
\email{marciano@fudan.edu.cn}
\affiliation{Department of Physics \& Center for Field Theory and Particle Physics, Fudan University, 200433 Shanghai, China}

\begin{abstract}
\noindent
We discuss how inflation can emerge from a four-fermion interaction induced by torsion. Inflation can arise from coupling torsion to  Standard Model fermions, without any need of introducing new scalar particles beyond the Standard Model. Within this picture, the inflaton field can be a composite field of the SM-particles and arises from a Nambu-Jona-Lasinio mechanism in curved space-time, non-minimally coupled with the Ricci scalar. The model we specify predicts small value of the r-parameter, namely $r\sim 10^{-4} \div 10^{-2}$, which nonetheless would be detectable by the next generation of experiments, including BICEP 3 and the AliCPT projects. 
\end{abstract}

\maketitle

Over the last 50 years, after the discovery of the Cosmic Microwave Background (CMB) by Penzias and Wilson, our understanding of cosmology and cosmography has undergone a terrific increase. The current picture of the Universe, which accounts for the flatness of its spatial hypersurfaces and the homogeneity of the CMB celestial distribution, highly favors the hypothesis that the Universe underwent an inflationary epoch in the early cosmological time, as confirmed recently by data from the Planck satellite \cite{Planck}. The origin of the inflaton field, mediating the inflation dynamics, is still unknown: an infinite class of Standard Model extensions or of extended theories of gravity are compatible with current CMB data. This is why in recent years an ``inflation'' of inflationary models has been suggested in literature. 

On the other hand, the minimalistic attitude, inspired by the {\it Occam's razor} principle, suggests to consider the possibility that the inflationary stage was not triggered by any new particle field beyond the Standard Model. In other words, the possibility that inflation was not mediated by any new inflaton field, introduced {\it ad hoc} or belonging to some Standard Model symmetry extensions, is aesthetically attractive as well as theoretically challenging. Of course, the Occam's razor principle cannot be taken literally, and should not prevent from considering the eventuality of physics beyond the Standard Model, which can only be falsified experimentally. 

We propose that inflation is triggered by the coupling of Standard Model fermions with gravitational torsion. Such a new mechanism can be realized within the context of the Einstein-Cartan-Holst-Sciama-Kibble theory (ECHSK) \cite{Torsion,t1,t2,t3,t4,t5,t6,t7,t8,t9,Torsion2,Torsion3,Perez:2005pm,Frei,Mag}. Here gravity is not extended to account for any extra degrees of freedom: any new higher-derivative terms or new field coupled to the rank-2 metric tensor are added to the Einstein-Hilbert Lagrangian in this framework. Within the ECHSK model, the Einstein-Hilbert kinetic term just recasts in terms of the spin-connection field and the vierbein. Internal consistency of the ECHSK theory requires a new coupling between the torsion component of the connection to the Standard Model fermion fields. If no other dynamical terms involving torsion are added to the Einstein-Hilbert fields, once these have been recast in terms of the spin-connection field, torsion will not be dynamical but will behave just as an auxiliary field. 

It is crucial to emphasize that, rather than a mere reformulation of effective models of inflation with put-in-by-hand fields, our idea provides a conceptual breakthrough in the physics of inflation. Up to now, most inflationary theories introduce the inflation field in an ad hoc way. What we will show below is not the replacement of one hypothetical field by another, but instead the realization that the put-in-by-hand inflaton field may actually be the effective description of an emergent phenomenon from fermion condensation. Furthermore, we point out that our notion can be distinguished from the put-in-by-hand inflaton scenario by means of the phenomenological analysis that we provide, which includes the prediction of a small tensor-to-scalar ratio $r$ that can be experimentally falsified. At the same time, we admit that our theory is an effective one, and thus cannot address neither the UV nor the trans-Planckian \cite{Staro1, Staro2, Martin:2000xs} problems. In stead, only the future development of a final theory of quantum gravity will be able to address properly these issues.

Integrating out torsion entails new effective four-fermion interactions, compatible with the Standard Model gauge symmetries. As a consequence, a sort of effective Nambu-Jona-Lasinio (NJL) mechanism emerges --- it is needless to remark that the dynamics is very different from the one in QCD, originally studied in \cite{NJL,NJL2}. Depending on the sign of the fermion-torsion coupling, the interaction can be either attractive or repulsive, due to the way energy conditions are fulfilled. When the phase is repulsive, possible bouncing cosmology scenarios~\footnote{In particular, in \cite{Lucat:2015rla} the Schwinger-Keldysh formalism has been adopted in order to describe evolution of bilinear condensates.} have been investigated in Refs.~\cite{Bambi:2014uua,Alexander:2014eva,Alexander:2014uaa,Lucat:2015rla,Addazi:2016rnz,Dona:2016fip}. Here we will instead show how inflation can be attained, thanks to similar mechanism, but exploring a different region of model parameters's space. Based on the ECHSK theory, we will obtain the effective interaction potential driving the NJL condensates. We will explore the hypothesis that such an effective potential triggers inflation and constrain its region of parameters from Planck data. We will show that this model predicts a scalar-to-tensor ratio in between the range $r\sim 10^{-4}\div 10^{-2}$, realized within a natural subset of the parameters's space. This prediction can be falsified by the next generation of experiments measuring B-mode polarizations, which include the BICEP 3 and AliCPT projects.

We can cast the Einstein-Cartan-Holst-Sciama-Kibble theory (ECHSK) in the first order formalism as follows: 
\be \label{DL}
S=\frac{1}{2\kappa}\int d^{4}x |e|e_{I}^{\mu}e_{J}^{\nu}P^{IJ}_{KL}F_{\mu\nu}^{KL}(\omega)
\,,
\ee
where 
$$F_{\mu\nu}^{IJ}=d\omega^{IJ}+\omega^{IL}\wedge \omega_{L}^{\ J}$$
is the field-curvature of $\omega^{IJ}$, the coupling constant $\kappa=8\pi G_{N}$ and 
$$P_{KL}^{IJ}=\delta_{K}^{[I}\delta_{L}^{J]}-\frac{1}{2\gamma}\epsilon_{KL}^{IJ}\,$$
contains the Levi-Civita symbol $\epsilon_{IJKL}$ multiplying the Barbero-Immirzi parameter $\gamma$. Thanks to this reformulation of GR, one can couple the spin-connection $\omega^{IJ}$ to Standard model fermions by means of
\be \label{DL2}
S_{\Psi}=\frac{1}{4}\int d^{4}x|e|\left[\imath\bar{\Psi}\gamma^{I}e_{I}^{\mu}\left(1-\frac{\imath}{\alpha}\gamma_{5}\right)\nabla_{\mu}\Psi \right]+h.c.
\,,
\ee
where $\alpha$ is a coupling constant. Within this action, the covariant derivative can be divided in a torsionless and a torsionful part. The torsional term induces every possible four fermion terms of the form 
\be \label{DL3}
S_{eff}=-\xi \kappa \int d^{4}x|e|J_{5}^{L}J_{5}^{M}\eta_{LM}\,,
\ee
where $J_{5}^{L}$ stand for the fermionic axial currents $J_{5}^{L}=\bar{\Psi}\gamma^{5}\gamma^{L}\Psi$ and $\xi$ is a combination of the microscopic couplings of the original Lagrangian $\alpha$ and $\gamma$ of the form
$$\xi=\frac{3}{16}\frac{\gamma^{2}}{1+\gamma^{2}}\left(1+\frac{2}{\alpha \gamma}-\frac{1}{\alpha^{2}}\right)\,.$$

In full generality, we obtain every possible four-fermion coupling compatible with the SM gauge group. All possible neutral quark and lepton axial currents are mixed with each other. For example, in the quark sector, four-fermion interaction like 
$$(\bar{q}^{a}_{l}\gamma^{L}\gamma_{5}q_{l\,a})(\bar{q}^{b}_{l}\gamma_{L}\gamma_{5}q_{l\,b}),\,\,\,(\bar{q}_{l}^{a}\gamma^{L}\gamma_{5}q_{l\,c})(\bar{q}_{l}^{b}\gamma_{L}\gamma_{5}q_{l\, d})\epsilon_{ab}\epsilon^{cd}$$
are generated from the torsion coupling. Analogous terms are generated in the leptonic sector. Finally, baryon/lepton conserving neutral mixing currents among quarks and leptons are sourced by torsion. The latter terms do not generate composite scalarons, but can introduce scalaron mixing. 

Generically, a quantum field theory analysis of this model that would take into account also loop corrections, would be very complicated to achieve --- see e.g. the models based on the potential derived by Coleman and Weinberg in Ref.~\cite{Coleman:1973jx}, and further improvements. Nonetheless, since the number of species are involving a large number of SM fields, we may perform the  large-N approximation \cite{Giddings:1992ae,Inagaki:1993ya,O2}. Thus, we will treat the full problem in a semiclassical $1/N$ framework, in order to keep under control the loop-corrections and the effective potential we will derive below. 

Moving from Eq.(\ref{DL}), the effective four-fermion action casts in the large N-approximation as
\begin{eqnarray} \label{DL4}
&&\int \sqrt{-g}d^{4}x\bar{\Psi} (\imath\gamma^{\mu}(x)\nabla_{\mu}-M)\Psi \\
&&+\frac{\lambda}{2N_{f}}[
(\bar{\Psi}\Psi)(\bar{\Psi}\Psi)+(\bar{\Psi}\imath\gamma_{5}\Psi)(\bar{\Psi}\imath\gamma_{5}\Psi)]\,, \nonumber
\end{eqnarray}
where $N_{f}$ is the number of fermions, $M$ is the fermion mass matrix, the coupling constant recasts $\lambda=\xi \kappa$,  the notation $\gamma^\mu(x)=e^\mu_I(x)\, \gamma^I$ is adopted and the interaction terms $(\bar{\Psi}\Psi)(\bar{\Psi}\Psi)$ and  $(\bar{\Psi}\imath \gamma_{5}\Psi)(\bar{\Psi}\imath \gamma_{5}\Psi)$ include all the  possible neutral charge four-fermion operators, compatible with the SM gauge symmetries. We may neglect the fermion masses in the following analysis, motivated by the high hierarchy among the SM particles masses and the inflation scale. We also neglect contributions arising from vector condensates.

The effective field theory of composite scalarons can be conveniently studied within the framework of the functional methods, by introducing auxiliary fields $\Pi$. The total action can then be recast as
\be \label{DL5}
S=S_{EH}+S_{\Pi}\, ,
\ee
where $S_{\Pi}$ is
\begin{eqnarray}  \label{DL5b}
&&\int \sqrt{-g}\, d^{4}x\,  [ \bar{\Psi} \imath \gamma^{\mu}(x)\nabla_{\mu}\Psi \\
&&-\frac{N_{f}}{2\lambda}(|\Pi|^{2}+|\Sigma|^{2})-\bar{\Psi} (\Sigma+\imath \gamma_{5}\Pi) \Psi ]\, . \nonumber
\end{eqnarray}
$\Pi$ and $\Sigma$ are matrices of scalar and pseudoscalar fields. 

The generating functional is 
\be \label{DL6}
Z[\eta,\bar{\eta}]=\int \mathcal{D}\Psi\,  \mathcal{D}\bar{\Psi} \, \mathcal{D}\Pi \, \mathcal{D}\Sigma \, \, e^{\imath S+\imath \bar{\eta}\Psi+\imath \bar{\Psi}\eta}\,,
\ee
where $\eta,\bar{\eta}$ are grassmannian source functions. Performing the grassmannian integration over all the fermions and setting the sources to zero, we obtain the effective partition functional 
\be \label{DL7}
Z[0,0]=\int \mathcal{D}\Pi \mathcal{D}\Sigma \, e^{\imath N_{f}S_{\rm eff}} \, ,
\ee
$S_{\rm eff}$ denoting 
\begin{eqnarray}
S_{\rm eff}=  &&\int d^{4}x \sqrt{-g}\{-\frac{1}{\lambda}(|\Pi|^{2}+|\Sigma|^{2}) \nonumber \\
&&-\imath \, {\rm ln\,Det}\{\imath I\gamma^{\mu}(x)\nabla_{\mu}-(\Sigma+\imath \gamma_{5}\Pi)\}  \}\, ,\nonumber 
\end{eqnarray}
where $I$ is a $N_{f}\times N_{f}$ identity matrix. The effective action receives negligible corrections controlled by the number of fermions, {\it i.e.} 
$$S_{\rm eff}[\Pi,\Sigma]+O(1/N_{f})\,.$$

From Eq.(\ref{DL7}), we obtain the effective matrix potential
\begin{eqnarray} \label{VPOT}
V(\Pi)=&&\frac{1}{2\lambda}(|\Pi|^{2}+|\Sigma|^{2}) \\
&&+\imath {\rm Tr\,ln}\langle x|\imath \gamma^{\mu}(x)I\nabla_{\mu}-(\Sigma+\imath \gamma_{5}\Pi)|z\rangle\,, \nonumber
\end{eqnarray}
where $\Pi$ and $\Sigma$ are treated as classical slow-varying fields. 

The second formal term of the expression can be estimated using the proper time method \cite{SC}. One obtains the following formal expression for the effective potential 
\be \label{VPOT2}
V=\frac{1}{2\lambda}(|\Pi|^{2}+|\Sigma|^{2})-\imath{\rm Tr\,ln}\,S(x,x,A)\,,
\ee
where $A=\Sigma+\imath \gamma_{5}\Pi$ and 
\be \label{S1}
S(x,y;A)=\langle x|(\imath I\gamma^{\mu}\nabla_{\mu}-A)^{-1}|y\rangle
\ee
is the matrix propagator associated to the classical matrix 
equation
\be \label{E1}
(\imath I \gamma^{\mu}(x)\nabla_{\mu}-A)S(x,y;A)=I\frac{1}{\sqrt{-g(x)}}\delta^{4}(x-y)\,.
\ee
Let us expand around the background 
$A=\bar{A}+\delta A$. 
\be \label{Det}
{\rm ln\,Det}\left\{\imath I\gamma^{\mu}(x)\nabla_{\mu}-A\right\}={\rm Tr\,ln}\left\{ \imath  I \gamma^{\mu}\nabla_{\mu}-A\right\}
\ee
$$={\rm Tr\,ln}\{ \imath I \gamma^{\mu}(x)\nabla_{\mu}-A\}-\int d^{4}{\rm Tr}\{\delta A(x)\,S_{F}(x,x)\}$$
$$-\frac{1}{2}\int d^{4}x\int d^{4}y \, \delta A(x) \, S_{F}(x,y)\, \delta A(y) \, S_{F}(y,x)+...\,,$$
where $S_{F}$ is the fermion propagator
provided by 
\be \label{SF}
\sqrt{-g}(\imath I \gamma^{\mu}(x)\nabla_{\mu}-M)S_{F}(x,y)=\imath \delta^{4}(x-y)I\,.
\ee

In large the $N$ and weakly varying curvature approximations, a general expression for the effective potential in curved space-time can be recovered. In particular, the propagator from x to x, corresponding to a bubble diagram of the effective scalar, is
\begin{eqnarray} \label{Sxx}
S(x,x;A)=&&\int \frac{d^{4}q}{(2\pi)^{4}}\Big[(I\gamma^{a}q_{a}+A)\frac{1}{q^{2}-|A|^{2}} \\
&&-\frac{1}{12}R(I\gamma^{a}q_{a}+A)\frac{1}{(q^{2}-|A|^{2})^{2}}\nonumber\\
&&+\frac{2}{3}R_{\mu\nu}q^{\mu}q^{\nu}(I\gamma^{a}q_{a}+A) 
\frac{1}{(q^{2}-|A|^{2})^{3}} \nonumber\\
&&-\frac{1}{8}\gamma^{a}[\gamma^{c},\gamma^{d}]R_{cda\mu}q^{\mu}\frac{1}{(q^{2}-|A|^{2})^{2}}\Big]\,. \nonumber
\end{eqnarray}

Within the weakly varying curvature approximation --- $\dot{R}\simeq 0$, compatibly with the inflationary regime --- we obtain the final effective potential $V(A)$ for the composite particles to be
\begin{eqnarray}
&V(A)\!=\!\tilde{V}(A)-\frac{1}{(4\pi)^{2}}\frac{R}{6}\left[-|A|^{2}{\rm ln}\left(1+\frac{\Lambda^{2}}{|A|^{2}}\right)+\frac{\Lambda^{2}|A|^{2}}{\Lambda^{2}+|A|^{2}}\right], \nonumber
\end{eqnarray}
where 
\begin{eqnarray}
&\tilde{V}=V_{0}+\frac{1}{2\lambda}|A|^{2} \nonumber\\
&-\frac{1}{4\pi^{2}}\left[|A|^{2}\Lambda^{2}+\Lambda^{4}{\rm ln}\left(1+\frac{|A|^{2}}{\Lambda^{2}} \right)-|A|^{4}{\rm ln}\left(1+\frac{\Lambda^{2}}{|A|^{2}} \right)\right]\,, \nonumber
\end{eqnarray}
$V_{0}=V(A)|_{A=0}$ and $\Lambda^{2}=c(\xi \kappa)^{-1}$ is the UV cutoff scale ($c$ is a numerical prefactor). The last term can be seen as $\omega(A)R$ term. In other words, this theory is a {\it composite multi scalars-tensor theory}. For every flavor of fermions $N_{f}$, we have $N_{f}$ composite states, {\it i.e.} $\Pi$ is a $N_{f}$ dimensional scalar multiplet. In particular, this expression can be instantiated within FLRW background, entailing 
\begin{eqnarray} \label{effecp}
&V(A)=V_{0}+\frac{1}{2\lambda}|A|^{2}  \nonumber \\
&-\frac{1}{4\pi^{2}}\left[|A|^{2}\Lambda^{2}+\Lambda^{4}{\rm ln}\left(1+\frac{|A| ^{2}}{\Lambda^{2}} \right)-|A| ^{4}{\rm ln}\left(1+\frac{\Lambda^{2}}{|A| ^{2}} \right)\right]  \nonumber \\
&-\frac{1}{(4\pi)^{2}}(\dot{H}+2H^{2})\left[-|A|^{2}{\rm ln}\left(1+\frac{\Lambda^{2}}{|A|^{2}}\right)+\frac{\Lambda^{2}|A|^{2}}{\Lambda^{2}+|A|^{2}} \right]\,, \nonumber 
\end{eqnarray}
where $V_{0}=V(A)|_{A=0}$. We will study now the effective potential in the slow-roll regime $\dot{H}<\!\!<H^{2}$. 

The potential Eq.(\ref{effecp}) involves mixing terms among all the scalar and pseudo-scalar fields belonging to the matrix multiplets $\Sigma$ and $\Pi$. This certainly leads to a highly complicated dynamics, typical of a multi-field scenario. Nonetheless, the situation is very much simplified if we consider an initial {\it custodial} global symmetry which is dynamically broken by the curvaton mechanism. In this case, a sub-group among all initial scalars and pseudo-scalars emerge as a pseudo Nambu-Goldstone bosons of the initial custodial symmetry, {\it i.e.} these fields are much lighter then the others. This is very much the same of what happens in QCD, where the pions and the $\eta$-mesons are pseudo-NG bosons of the chiral symmetry, while the other mesons are much more massive --- like the $\eta'$-meson. Another possibility is to select the possible four-fermion interactions by imposing {\it flavor} or {\it horizontal} gauge symmetries --- similarly to what was done within the context of large extra-dimensions scenarios, in order to avoid dangerous flavor changing neutral currents \cite{Berezhiani:1998wt}.

Thus, we can suppose that only these pseudo-NB bosons will trigger inflation, the other ones being more massive. In the simpler case, we can suppose the the curvaton mechanism dynamically breaks an initial vector or axial global $U(1)$.  Then only one (pseudo-)scalaron remains lighter than the other ones, and a single inflation scenario can be envisaged. Within the case of an initial vector-like global symmetry $U(1)_{V}$, we finally obtain the single field potential
\begin{eqnarray} \label{effecps}
&V(a)=V_{0}+\frac{1}{2\lambda}|a|^{2} \nonumber\\
&-\frac{1}{4\pi^{2}}\left[|a|^{2}\Lambda^{2}+\Lambda^{4}{\rm ln}\left(1+\frac{|a| ^{2}}{\Lambda^{2}} \right)-|a| ^{4}{\rm ln}\left(1+\frac{\Lambda^{2}}{|a| ^{2}} \right)\right] \nonumber\\
&\!\!\!\!\!\!\!\!-\frac{1}{(4\pi)^{2}}(\dot{H}+2H^{2})\left[-|a|^{2}{\rm ln}\left(1+\frac{\Lambda^{2}}{|a|^{2}}\right)+\frac{\Lambda^{2}|a|^{2}} {\Lambda^{2}+|a|^{2}} \right]\!\!, 
\end{eqnarray}
where $a$ is the pseudo-NG boson of the initial $U(1)_{V}$. 


\begin{figure}[t]
\centerline{ \includegraphics [height=6cm,width=0.95\columnwidth]{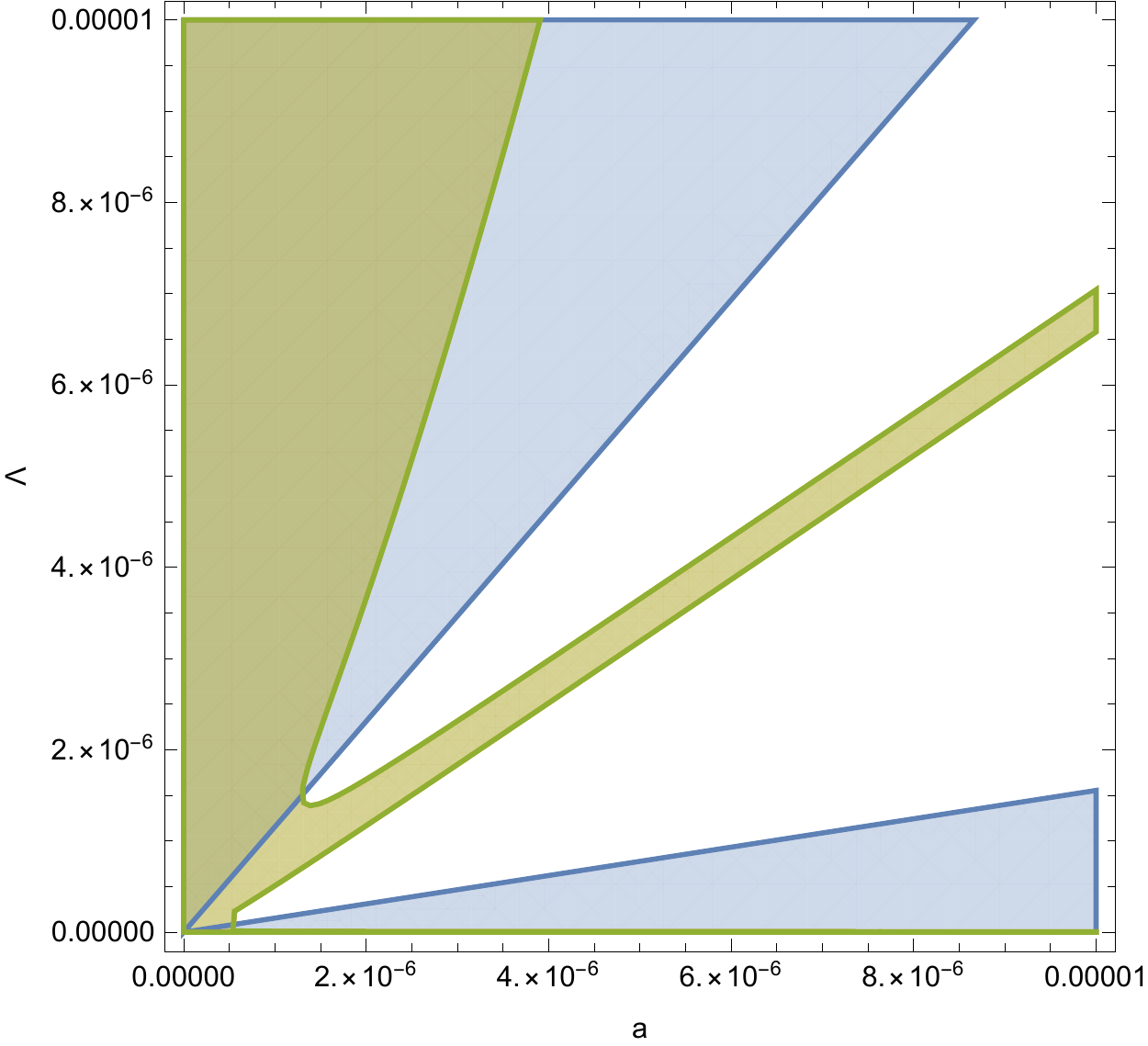}}
\vspace*{-1ex}
\caption{Considering the single-field inflation case, we display in units of Planck mass the $(\Lambda,\ a)$ space that is allowed by the constraints on $n_{s}$ and $\Delta_{R}$ that arise from the Planck satellite data. Outside the blue and green subregions values of $\Lambda$ and $a$ are compatible with the Planck data. }
\label{plot}   
\end{figure}


We can now discuss the phenomenology of the effective composite scalars emerging from the torsion coupling. In particular, we will show that the effective composite scalar may provide a good candidate for inflation.  As is known, the slow roll parameters $\epsilon,\eta$ are related to the effective inflaton potential Eq.~(\ref{effecps}) as 
\be \label{Pi1}
\frac{\epsilon[a]}{M_{Pl}^{2}}=\frac{1}{2}\left(\frac{V'[a]}{V[a]}\right)^{2},\,\,\,\frac{\eta[a]}{M_{Pl}^{2}}=\frac{V''[a]}{V[a]}\,.
\ee

We can put severe constraints on the parameter spaces of our model allowed by the observational data. 
The $\epsilon$-parameter is related to 
\be \label{Delta}
\Delta_{R}^{2}\simeq \frac{V_{0}}{24\pi^{2}M_{Pl}^{4}\epsilon}\,,
\ee
and is constrained to be 
\be \label{Deltaexp}
\Delta_{R, \rm \, exp}^{2}=2,215\times 10^{-9}
\ee
from the Planck data \cite{Planck}. On the other hand, taking into account the observed value of the spectral index 
\be \label{n}
n_{s, \rm exp}= 0.968 \pm 0.006\,,
\ee
the number of e-folding, which is expressed by the relation 
\be \label{N}
N=\frac{1}{M_{Pl}^{2}}\int_{\phi_{end}}^{\phi}d\phi \frac{V}{V'} \,,
\ee
is constrained to be approximately $60$, corresponding to $\epsilon \simeq 0.003$ and $|\eta|\simeq 0.02$ --- the explicit $N$-dependence of the slow-roll parameters reads $\epsilon={1}/{(2N/3+1)^{3/2}}$ and $\eta=-{1}/{(2N/3+1)}$, from which is inferred the $N$-dependence of the spectral index considering the relation $n_{s}-1= - 2\, \epsilon - \, \eta$.

Relying on these constraints, we are able to exclude a large subspace of parameters. First, we find that the model is compatible with the Planck data if $\lambda^{-1} = \Lambda^{2}/2\pi^{2}$, for any choice of $\Lambda, H$. In Fig.~1, we show the constraints to the single inflaton field configuration and the UV scale $\Lambda$ arising from $n_{s, \rm exp}$ and $\Delta^2_{R, \, \rm exp}$. In particular, we find that, for $\Lambda=10^{-3}M_{Pl}$, $V_{0}=\Lambda^{4}$, Planck constraints can be easily satisfied. The excursus of the inflaton field during inflation approaches the Planck scale without exceeding it. We also point out the existence of another {\it critical branch}, $\lambda^{-1}=\Lambda^{2}/\pi^{2}$, which is subtly compatible with the Planck data only for $|a|\!<\!\!<\!\Lambda$. Within this regime, the third line of Eq.~\eqref{effecps} provides a quadratic term for the potential, that together with a small quartic interaction arising from the first two lines of Eq.~\eqref{effecps}, ensures compatibility with the Planck data. Furthermore, for $|a|\!<\!\!<\!\Lambda$, once the scalar field approaches the bottom of the potential, a perturbative reheating phase is approached, in which the potential converges to the form of the Coleman-Weinberg potential. This ensures a graceful exit mechanism from inflation for this model, with a reliable reheating mechanism \cite{Cerioni:2009kn,Barenboim:2013wra}.
 
We finally remark that for the low energies involved in this effective description, our analysis is consistent with other regularization schemes, such as the ones described in Refs.~\cite{CR,Giacosa:2008rw,MW}.

As is well-known, the $r$ parameter is defined as 
$$r=\frac{\mathcal{P}_{T}(k_{*})}{\mathcal{P}_{\zeta}(k_{*})}\,,$$
where $k^{*}$ is the so-called pivot scale ($k_{*}=0.002\, Mpc^{-1}$). For $\Lambda \simeq 2\times 10^{15}\div 10^{16}\, {\rm GeV}$, the tensor spectrum casts  
$$\mathcal{P}_{T}=\frac{2H_{inf}^{2}}{\pi^{2}M_{Pl}^{2}}\simeq \frac{2V_{0}}{3\pi M_{Pl}^{4}}
\sim 10^{-13}\div 10^{-11}\,,$$
where $H_{\rm inf}$ is the Hubble expansion rate during inflation. The scalar spectrum must be $\mathcal{P}_{\zeta}\sim 2.1 \times 10^{-9}$, implying $r=10^{-4}\div 10^{-2}$. As a consequence, for a UV scale close to the GUT scale, our model predicts a $r$ parameter value that is detectable in next generation of experiments like the BICEP 3 and ALI projects.

In conclusion, we have explored an inflation mechanism that is originated from the torsion-fermion coupling, within the context of the  Einstein-Cartan-Holst-Sciama-Kibble theory. In particular, we have shown how the torsion induces effective four-fermions interaction
and how the low-energy dynamics can be described from an effective Nambu-Jona-Lasinio model. We have computed the effective interaction potential driving the NJL fields, and we have compared it with current constraints arising from Planck data. We have put stringent constraints on the parameters space of the model, and have shown that the model is not ruled out if and only if the four-fermion coupling $\lambda$ generated by the torsion have a precise critical value. 

We have discussed how the r-parameter can be high as $r\sim 10^{-2}$, in a natural sub-region of possible parameters
entering the effective potential. This scenario can be falsified by B-mode phenomenology that will be developed by forthcoming data from the BICEP 3 and ALI-CMB collaborations.   

Finally, we comment on the case of of inflation with multi-composite fields of the NJL model --- which was not analyzed in this paper. 
We have limited our-self to the phenomenological analysis of the single-field inflation, invoking a custodial or flavor symmetries.
However, a multi-field inflation scenario can be naturally envisaged within this context, leading to non-gaussianities in the CMB. 
This case lies in the effective field theory parametrization for multi-field inflation proposed in Ref.~\cite{Senatore:2010wk}. Further constraints will arise in this scenario from the analysis of cross-correlation functions, which are naturally generated in the multi-composite fields fermionic approach, the latter accounting for parity violating terms as well.

\vspace{1cm}
{\it{\textbf{Acknowledgement.}}}\\
\noindent
We wish to thank Misao Sasaki for enlightening discussions. 
A.A.\ and A.M.\ acknowledge support by the NSFC, through the grant No. 11875113, the Shanghai Municipality, through the grant No. KBH1512299, and by Fudan University, through the grant No. JJH1512105.

\end{document}